\documentclass[newstyle,twocolumn,proceedings]{rmaa}

\title{A New Approach in Data Reduction: Proper Handling of Random Errors and
Image Distortions}
\author{N.~Cardiel, J.~Gorgas, J.~Gallego, \'{A}.~Serrano, J.~Zamorano,
        \affil{Departamento de Astrof\'{\i}sica,
               Universidad Complutense de Madrid, Spain}
        M.~L.~Garc\'{\i}a-Vargas, P.~G\'{o}mez-Cambronero, and J.~M.~Filgueira
        \affil{Gran Telescopio Canarias, La Laguna, Spain}}
\fulladdresses{%
\item N.~Cardiel, J.~Gorgas, J.~Gallego, \'{A}.~Serrano, and J.~Zamorano:
      Departamento de Astrof\'{\i}sica, Facultad de F\'{\i}sicas,
      Universidad Complutense de Madrid, 28040~Madrid, 
      Spain (ncl@astrax.fis.ucm.es).
\item M.~L.~Garc\'{\i}a-Vargas, P.~G\'{o}mez-Cambronero, and J.~M.~Filgueira:
      Gran Telescopio Canarias (GTC), C/ V\'{\i}a L\'{a}ctea s/n
      (Instituto de Astrof\'{\i}sica de Canarias),
      38200~La~Laguna (Tenerife), Spain.}

\shortauthor{Cardiel~et~al.}
\shorttitle{A New Approach in Data Reduction}


\resumen{%
Los procesos de reducci\'{o}n de datos tienen como objetivo minimizar el
impacto que las imperfecciones en la adquisici\'{o}n de los mismos producen en
la obtenci\'{o}n de medidas de inter\'{e}s para el astr\'{o}nomo. Para
conseguir este objetivo, es necesario realizar manipulaciones aritm\'{e}ticas,
utilizando im\'{a}genes de datos y de calibraci\'{o}n.  Por otro lado, la
interpretaci\'{o}n correcta de las medidas s\'{o}lo es posible cuando existe
una determinaci\'{o}n precisa de los errores asociados.  En este trabajo
discutimos diferentes estrategias posibles para obtener determinaciones
realistas de los errores aleatorios finales. En concreto, destacamos los
beneficios que conlleva considerar el proceso de reducci\'{o}n de datos como la
caracterizaci\'{o}n completa de las im\'{a}genes originales, pero evitando,
tanto como sea posible, la alteraci\'{o}n aritm\'{e}tica de las im\'{a}genes
hasta el momento de su an\'{a}lisis final y obtenci\'{o}n de medidas
definitivas.  Esta filosof\'{\i}a de reducci\'{o}n ser\'{a} utilizada en la
reducci\'{o}n de datos de ELMER y de EMIR.}

\abstract{%
Data reduction procedures are aimed to minimize the impact of data acquisition
imperfections on the measurement of data properties with a scientific meaning
for the astronomer. To achieve this purpose, appropriate
arithmetic manipulations with data and calibration frames must be performed.
Furthermore, a full understanding of all the possible measurements relies 
on a solid constraint of their associated errors. We discuss different
strategies for obtaining realistic determinations of final random errors.
In particular, we highlight the benefits of considering the data
reduction process as the full characterization of the raw-data frames, 
but avoiding, as far as possible, the arithmetic manipulation of that data 
until the final measure and analysis of the image properties. This philosophy
will be used in the pipeline data reduction for ELMER and EMIR.}

\listofauthors{N.~Cardiel, J.~Gorgas, J.~Gallego, \'{A}.~Serrano, J.~Zamorano,
               M.~L.~Garc\'{\i}a-Vargas, P.~G\'{o}mez-Cambronero \&
               J.~M.~Filgueira}
\indexauthor{Cardiel, N.}
\indexauthor{Gorgas, J.}
\indexauthor{Gallego, J.}
\indexauthor{Serrano, \'{A}.}
\indexauthor{Zamorano, J.}
\indexauthor{Garc\'{\i}a-Vargas, M.~L.}
\indexauthor{G\'{o}mez-Cambronero, P.}
\indexauthor{Filgueira, J.~M.}

\begin{document}
\maketitle

\section{Introduction}

The Gran Telescopio Canarias (GTC)\footnote{http://www.gtc.iac.es}, as one the
best human tools to explore and reveal the unknown Universe, will give access,
in conjunction with its pioneering instrumentation, to rather faint and/or
distant objects, in practice inaccesible for 4~m class telescopes. For that
reason, very high signal-to-noise ratios are expected to be uncommon in most
cases. Under these circumstances, an accurate error estimation is essential to
guarantee the reliability of the measurements.

Although there are no magic recipes to quantify systematic errors in a general
situation, where a case by case solution must be sought, the state is,
fortunately, not so bad concerning random errors. Initially, the latter can be
measured and properly handled using typical statistical tools. In this
contribution we discuss the benefits and drawbacks of different methods to
quantify random errors in the context of data reduction pipelines. After
examining the possibilities, we conclude that the classic reduction procedure
is not perfectly suited for error handling. In this sense, the
responsibility for the completion of the more complex data reduction steps must
be transferred to the analysis tools. For this approach to be possible,
additional information must also be provided to those tools, which in turn
implies that the reduction process should be modified in order to produce that
information. A discussion concerning the treatment of systematic errors is out
of the scope of this paper.

\begin{figure*}
\begin{center}
 \resizebox{0.8\hsize}{!}{%
 \includegraphics[angle=0]{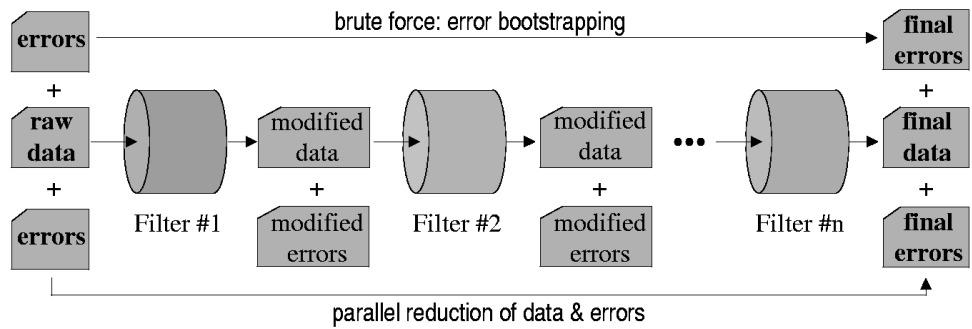}}
 \caption{Classic reduction procedure.}
 \label{figure_classic_reduction}
\end{center}
\end{figure*}

\section{The Classic Reduction Procedure}

\subsection{Three methods to quantify random errors}

In a classic view (see Figure~\ref{figure_classic_reduction}), a typical data
reduction pipeline can be considered as a collection of filters, each of which
transforms input images into new output images, after performing some kind of
arithmetic manipulation and making use of additional measurements and
calibration frames when required.  Under this picture, three different
approaches can in principle be employed to determine random errors in
completely reduced images:

{\it i) Comparison of independent repeated measurements.\/} This is one of the
simplest and most straightforward ways to estimate errors, since, in
practice, errors are not computed nor handled through the reduction procedure.
The only requirement is the availability of a non too small number of
independent measurements. Although as such can be considered even the flux
collected by each {\it independent\/} pixel in a detector (for example when 
determining the sky flux error in direct imaging), in most cases this method
requires the comparison of different frames. For that reason, and given
that for many purposes it may constitute an extremely expensive method in terms 
of observing time, its applicability on a general situation seems rather 
unlikely.

{\it ii) First principles and brute force: error bootstrapping.\/} Making use
of the knowledge concerning how photo-electrons are generated (expected
statistical distribution of photon arrival into each pixel, detector gain and
read-out noise), it is possible to generate an error image associated to each
raw-data frame. By means of error bootstrapping via Monte Carlo simulations,
new instances of the initial raw-data frame are simulated and can be completely
reduced as if they were real observations. The comparison of the 
measurements performed over the whole set of reduced simulated observations 
provides then a good estimation of the final errors. However, and although 
this method overcome the problem of wasting observing time, it can also be
terribly expensive, but now in terms of computing time.

{\it iii) First principles and elegance: parallel reduction of error and 
data frames.\/} Instead of wasting either observing or computing time,
it is also possible to feed the data
reduction pipeline with both, the original raw-data frame and its associated
error frame (computed from first principles), and proceed only once throughout
the whole reduction process. In this case every single arithmetic manipulation
performed over the data image must be translated, using the law of propagation
of errors, into parallel manipulations of the error image. Unfortunately,
typical astronomical data reduction packages (e.g.\ Iraf, Midas, etc.) 
do not consider random error propagation as a default operation and, thus,
some kind of additional programming is unavoidable.

\begin{figure*}
\begin{center}
 \resizebox{0.8\hsize}{!}{%
 \includegraphics[angle=0]{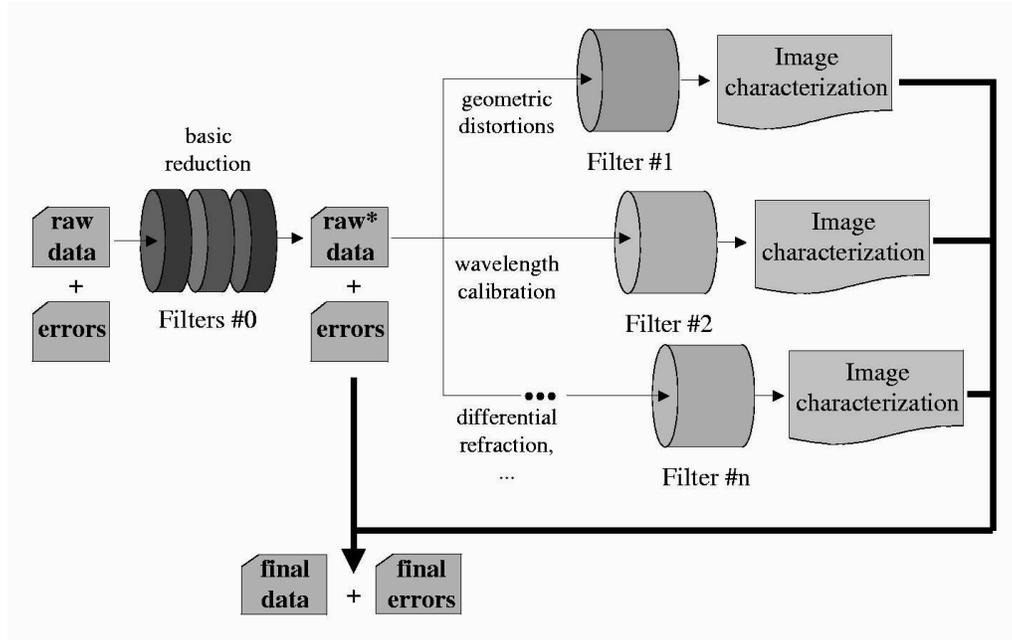}}
 \caption{Modified reduction procedure.}
 \label{figure_modified_reduction}
\end{center}
\end{figure*}

\subsection{Error correlation --- A real problem}

Although each of the three methods described above is suitable of being
employed in different circumstances, the third approach is undoubtedly the one
that, in practice, can be used in a more general situation. In fact, once the
appropriate data reduction tool is available, the parallel reduction of data
and error frames is the only way to proceed when observing or computing time
demands are prohibitively high. However, due to the unavoidable fact that the
information collected by detectors is physically sampled in pixels, this
approach collides with a major problem: errors start to be correlated as soon
as one introduces image manipulations involving rebinning or non-integer pixel
shifts of data. A naive use of the analysis tools would neglect the effect of
covariance terms, leading to dangerously underestimated final random errors.
Actually, this is likely the most common situation since, initially, the
classic reduction operates as a black box, unless specially modified for the
contrary. Unfortunately, as soon as one accumulates a few reduction steps
involving increment of correlation between adjacent pixels (e.g.\ image
rectification when correcting for geometric distortions, wavelength calibration
into a linear scale, etc.), the number of covariance terms starts to increase
too rapidly to make it feasible the possibility of stacking up and propagate
all the new coefficients for every single pixel of an image.

\begin{figure*}
\begin{center}
 \resizebox{0.8\hsize}{!}{%
 \includegraphics[angle=0]{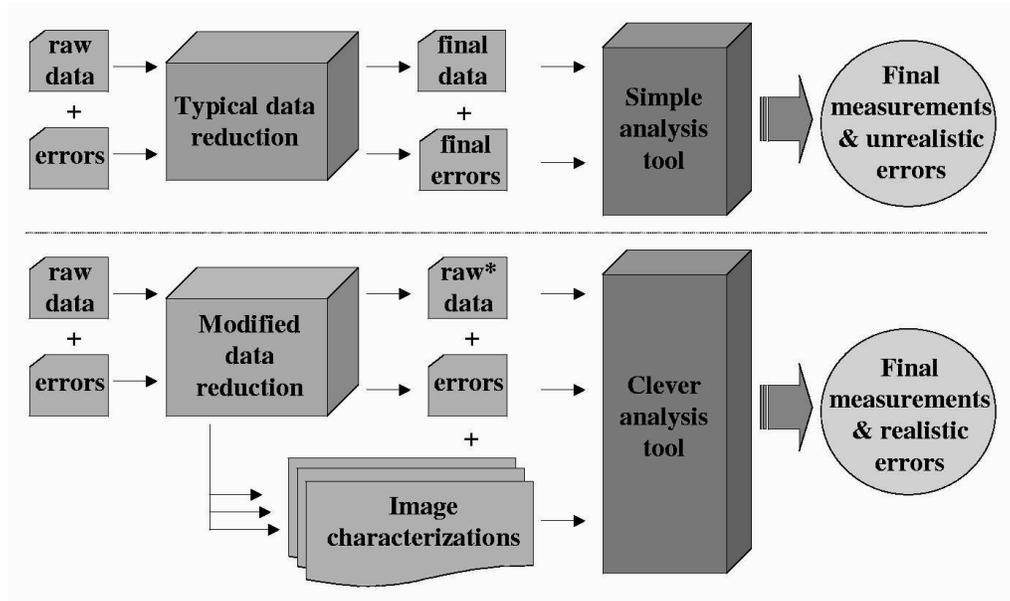}}
 \caption{Comparison between classic (upper panel) and modified (lower panel)
 reduction procedures.}
 \label{figure_modus_operandi}
\end{center}
\end{figure*}

\section{The Modified Reduction Procedure}

\subsection{Image Characterization}

Obviously, the problem can be circumvented if one prevents its emergence,
i.e.\ if one does not allow the data reduction process to introduce 
correlation into neighboring pixels before the final analysis. In other words,
if all the reduction steps that lead to error correlation are performed in a
single step during the measurement of the image properties with a scientific
meaning for the astronomer, there are no previous covariance terms to 
be concerned with. Whether this is actually possible or not may depend on the
type of reduction steps under consideration. In any case, a change in the
philosophy of the classic reduction procedure can greatly help in alleviating
the problem. The core of this change consists in considering the reductions
steps that originate pixel correlation as filters that 
{\it do not necessarily\/} take input images and
generate new versions of them after applying some kind of arithmetic
manipulation, but as filters that properly {\it characterize\/} the
image properties, without modifying those input images.

More precisely, the reduction steps can be segregated in two groups (see
Figure~\ref{figure_modified_reduction}): a) simple steps, which do not require
data rebinning nor non-integer pixel shifts of data; and b) complex steps,
those suitable of introducing error correlation between adjacent pixels. The
former may be operated like in a classic reductions, since their application do
not introduce covariance terms. However, the complex steps are only allowed to
determine the required image properties that one would need to actually perform
the correction. For the more common situations, this characterizations may be
simple polynomials (in order to model geometric distortions, non-linear
wavelength calibration scales, differential refraction dependence with
wavelength, etc.). Under this view, the end product of the modified reduction
procedure is constituted by a slightly modified version of the raw data frames
(after quite simple arithmetic manipulations) and by an associated collection
of image characterizations.

\subsection{Modus Operandi}

Clearly, at any moment it is possible to combine the result of the partial
reduction after all the linkable simple steps, with the information achieved
through all the characterizations derived from the complex steps, to obtain the
same result than in a classic data reduction (thick line in
Fig.~\ref{figure_modified_reduction}). However, instead of trying to obtain
completely reduced images ready for starting the analysis work, one can
directly feed a {\it clever analysis tool\/} with the end products of the
modified reduction procedure (see Figure~\ref{figure_modus_operandi}).
Obviously, this clever analysis tool has to perform its task taking into
account that some reductions steps have not been performed. For instance, if
one considers the study of a 2D spectroscopic image, the analysis tool should
use the information concerning geometric distortions, wavelength calibration
scale, differential refraction, etc., to obtain, for example, an equivalent
width through the measurement in the partially reduced (uncorrected for
geometric distortions, wavelength calibration, etc.) image. 
\adjustfinalcols
To accomplish this task, it is necessary to manipulate the data using a new and
distorted system of coordinates that must override the orthogonal coordinate
system defined by the physical pixels.  It is in this step where the final
error of the equivalent width should be obtained.  It is important to highlight
that, in this situation, such error estimation should not be a complex task,
since the analysis tool is supposed to be handling uncorrelated pixels.

The described reduction philosophy will be incorporated into the pipeline data
reduction for ELMER (http://www.gtc.iac.es/instrumentation/elmer\_s.asp) and
EMIR (http://www.ucm.es/info/emir).

\acknowledgements 

Financial support for this research has been provided by the Spanish Programa
Nacional de Astronom\'{\i}a y Astrof\'{\i}sica under grants AYA2000-977 and
AYA2000-1790.  This work has been benefitted by the experience of reducing
NIRSPEC data obtained at Keck~II. In this sense, N.C. acknowledges partial
financial support from a UCM Fundaci\'{o}n del Amo Fellowship, and a short
contract at UCSC.

\end{document}